\documentclass[english,oneside]{article}
\usepackage[T1]{fontenc}
\usepackage[latin1]{inputenc}
\usepackage{amsmath}
\usepackage{amssymb}
\usepackage[dvips]{epsfig}
\usepackage{comment}
\RequirePackage{setspace}

\newcommand{\halfo}{\frac{1}{2}}	%one half (half one)
\newcommand{\half}[1]{\frac{#1}{2}}
\newcommand{\ka}{\kappa}
\newcommand{\equ}[1]{\begin{equation} #1 \end{equation}}
\newcommand{\ba}{\begin{align}}
\newcommand{\ea}{\end{align}}	%doesn't work, don't know why
\newcommand{\eref}[1]{eq. (\ref{#1})}
\newcommand{\fref}[1]{fig. \ref{#1}}
\newcommand{\nnnl}{\nonumber\\}	%nonumber new line

\newcommand{\nq}{\nu_1}	%q is under 1
\newcommand{\nw}{\nu_2}	%w is under 2
\newcommand{\fig}[4]{\begin{figure}[#1]\centering\epsfig{file=#3}\caption{#4}\label{#2}\end{figure}}

\makeatletter

\newcommand{\lyxaddress}[1]{
\par {\center #1
\vspace{1.4em}
\noindent\par}
}

\usepackage{babel}
\makeatother
\begin{document}

\title{The infrared behavior of Landau gauge Yang-Mills theory \\ in $d$ = 2, 3 and 4 dimensions}

\author{Markus Huber$^{1}$, Reinhard Alkofer$^{1}$, Christian S. Fischer$^{2}$ and Kai Schwenzer$^{1}$}

\date{}

\maketitle

\begin{center}
\lyxaddress{$^{1}$Institut f\"ur Physik, Universit\"at Graz, Universit\"atsplatz 5,
8010 Graz, Austria\\
$^{2}$Institut f\"ur Kernphysik, Darmstadt University of Technology, \\
Schlossgartenstra\ss e 9, 64289 Darmstadt, Germany}
\end{center}

\begin{abstract}
We develop a general power counting scheme for the infrared limit
of Landau gauge $SU(N)$ Yang-Mills theory in arbitrary dimensions.
Employing a skeleton expansion, we find that the infrared behavior
is qualitatively independent of the spacetime dimension $d$. In the 
cases $d\!=\!2,3$ and $4$ even the quantitative results for the infrared 
exponents of the vertices 
differ only slightly. 
Therefore, corresponding 
lattice simulations provide interesting qualitative information for the 
physical case. We furthermore find that the loop integrals depend only 
weakly on the numerical values of the IR exponents.
\end{abstract}

\paragraph*{Introduction:}

The infrared (IR) regime of Yang-Mills theory determines the fundamental 
non-pertur\-bative properties of the non-Abelian gauge dynamics. 
In particular, the IR limit of Greens functions of the colored fields 
are likely connected to the problem of confinement as demonstrated within 
the scenarios of Kugo-Ojima \cite{Kugo:1979gm} and Gribov-Zwanziger 
\cite{Gribov:1977wm,Zwanziger:1991gz}. Due to the intricate nature of the 
problem an investigation via all available methods is desirable. Such 
combined efforts via Dyson-Schwinger equations (DSE) \cite{vonSmekal:1997is,Zwanziger:2001kw,Lerche:2002ep,Alkofer:2000wg}, renormalization group 
(RG) techniques \cite{Pawlowski:2003hq} and lattice gauge theory studies \cite{Sternbeck:2005tk} led during the
last years to a coherent picture of the infrared regime in Landau gauge. Hereby, in the continuum approaches, the IR scaling limit of general Greens functions in four spacetime dimensions has been determined \cite{Alkofer:2004it}. 
It exhibits, as argued for in \cite{Zwanziger:2003cf}, a dominance of the gauge fixing part of the action, i.e. the Faddeev-Popov ghost. 
A simultaneous analysis of DSE and RG methods allowed furthermore to show 
that this IR fixpoint is unique \cite{Fischer:2006vf}.

The Greens functions of Yang-Mills theory in Landau gauge are
studied extensively on the lattice, see e.g. \cite{Sternbeck:2005tk}. These 
analyses show the IR increase of the ghost dressing function required 
in both the Kugo-Ojima and the Gribov-Zwanziger scenario, as well as the IR 
scaling of the gluon dressing function in accordance with DSE
and RG analyses. The influence of finite volume corrections has been studied 
within DSE analyses on compact manifolds \cite{Fischer:2005ui} which agree qualitatively with the 
finite IR limit of current lattice results. However, a 
confirmation that even the gluon propagator vanishes weakly in the IR, 
which is necessary in the Gribov-Zwanziger scenario, has not yet been possible 
in four-dimensional lattice simulations due to the large lattice sizes required 
to probe the IR regime. The situation is similar for the important but 
even harder determination of the scaling behavior of vertex functions or the 
quantitative extraction of IR exponent. Lower dimensional lattices could 
allow to test qualitative aspects of these Greens functions in a substantially 
simpler setting. Naturally,
such a surrogate study requires that the generic features are identical to 
those in four-dimensional spacetime. Recent lattice studies in two \cite{Maas:2007uv} 
and three spacetime dimensions \cite{Cucchieri:2006tf} indeed suggest a power-law 
behavior similar to the four dimensional case.

$SU(N)$ Yang-Mills theory in three spacetime dimensions is also relevant
from two other points of view. In the high temperature limit only
the lowest Matsubara mode contributes, leading to a dimensional reduction
and thus to an effective field theory in three dimensions. The qualitative aspects
of this effective theory, in particular with regard to its confining
nature, may be covered by ordinary three-dimensional gauge theory \cite{Maas:2004se}. 
Similarly, Yang-Mills theory in the canonical quantization approach in Coulomb 
gauge \cite{Schleifenbaum:2006bq}  with an appropriate choice for the vacuum 
wave functional is structurally similar
to the corresponding three-dimensional theory in Landau gauge.

These results motivate to study the dependence of the infrared limit on the 
parameters of the theory in more detail. It will turn out that the infrared solution of 
Yang-Mills theory is determined by the spacetime 
dimension $d$ and one parameter $\kappa$ that determines the 
IR exponents of propagators and vertices. In this work we establish 
a manifest power counting scheme for general vertex functions of Landau gauge 
Yang-Mills theory in arbitrary dimensions. We find qualitatively similar results 
as in the four-dimensional analysis given in \cite{Alkofer:2004it}. Furthermore 
we discuss the dependence of the Dyson-Schwinger equations on the infrared scaling 
parameter $\kappa$. In four dimensions it takes a value $\kappa\approx0.595$ within 
an approximation based on the propagator DSEs and using 
%%%%%begin
%
%bare vertices \cite{Lerche:2002ep}. Other ansaetze for the vertices and in 
%particular their dynamical inclusion in the DSE system should change this 
%value. Therefore we study to what extend these dynamical building blocks in 
%the DSEs depend on $\kappa$, employing an analysis of the generic vertex 
%integrals. This allows to exclude possible alternative solutions of the DSE system.
a bare ghost-gluon vertex. 
A possible dressing of this vertex 
could change this value slightly \cite{Lerche:2002ep}. Therefore we study to 
what extent these dynamical building blocks in the DSEs depend on $\kappa$, 
employing an analysis of the IR-dominant integrals. 
%This allows to uniquely
%identify one branch of solutions for $\kappa$ as a function of dimension.
%%%%%end

\paragraph*{IR exponents for arbitrary $d$:}

In the following we will perform a scaling analysis for the IR regime of Yang-Mills 
theory in arbitrary dimensions. In contrast to the generic IR limit we study the 
limit where the coupling $g$ is kept fixed and has no inherent scaling dependence. 
Instead in dimensions other than four it fixes the fundamental scale of the theory. 
This limit is more directly accessible in lattice simulations and has been analyzed 
in \cite{Maas:2007uv,Cucchieri:2006tf}. 
A similar IR analysis has previously been performed for the physically important 
case $d\!=\!4$ in refs. \cite{Alkofer:2004it,Fischer:2006vf}. Although the scaling in arbitrary dimension may be abstracted from these results via the consideration of the appropriate canonical scaling dimensons, we chose to give a concise and nevertheless self-contained derivation of the general scaling relations and refer the reader to the corresponding work for details.
%We refer the reader to the corresponding
%work for details while we will here only sketch the
%main steps and point out the differences appearing in the general
%case. 
We note, that our analysis is immediately applicable to the gauge sector of 
QCD since closed quark loops are IR-suppressed due to the finite current quark 
masses \cite{Appelquist:1974tg} and therefore there are {\em no} quark contributions 
to purely bosonic Greens functions in the IR regime. This has been verified in ref. 
\cite{Fischer:2003rp}. \\
Analog to the corresponding analysis in four dimensions \cite{Alkofer:2004it},
the starting point for the IR analysis is the non-renormalization
of the ghost-gluon vertex in Landau gauge \cite{Taylor:1971ff}, 
%%%%%begin
%
which implies a finite vertex in the infrared \cite{Alkofer:2004it}. 
%%%%%end
This property depends purely on the transversality of the gluon propagator
and is, as long as the ghost-gluon scattering kernel is not strongly IR-divergent, 
therefore valid in arbitrary dimensions.
%%%%%begin
%
We come back to this point below.
%%%%%end
\fig{ht}{fig:gh-DSE-again}{./gh-DSE,width=12cm}{The DSE for the ghost propagator.} \\
%As shown before in the four dimensional case, the description of Yang-Mills theory in 
%terms of its fundamental gauge degrees of freedom does not break down in the infrared 
%regime \cite{Alkofer:2004it} even though these are no asymptotic physical states. For the moment we % assume that this holds in arbitrary dimension, too.
The infrared behavior of the propagators and vertices well below its inherent scale $g^{2/(4-d)}$ (respectively $\Lambda_{QCD}$ in $d=4$)
is determined via renormalization group arguments by scaling relations. The propagators 
of the gluons and ghosts 
\begin{align}
D_{\mu\nu}(p^2)&=\left( \delta_{\mu \nu}-\frac{p_\mu p_\nu}{p^2} \right) 
\frac{Z(p^2)}{p^2} \, , & D^G(p^2)&=-\frac{G(p^2)}{p^2} \, ,
\end{align}
are given in terms of dressing functions whose IR behavior is described 
by a power law ansatz
\begin{align}\label{eq:power-laws-props}
Z(p^2)&=c_{0,2} \cdot(p^2)^{\delta_{0,2}} \, , & G(p^2)&= c_{2,0} \cdot(p^2)^{\delta_{2,0}} \, ,
\end{align}
and similar for the vertices. Here we denote the IR exponent of 
a vertex with $2n$ ghost and $m$ gluon legs by $\delta_{2n,m}$ and the corresponding 
coefficient by $c_{2n,m}$. Whereas this coefficient is a constant for the propagators 
it is generally a function of $2n\!-\!m\!-\!1$ momentum ratios. \\
For the integral on the right hand side of the ghost propagator Dyson-Schwinger equation, 
cf. fig. \ref{fig:gh-DSE-again}, one can use the standard expression 
\cite{Lerche:2002ep,Anastasiou:1999ui} 
\begin{equation}
\label{eq:2-point-integral}
\int \frac{d^dq}{(2\pi)^d} (q^2)^{\nu_1}((q-p)^2)^{\nu_2}=(4\pi)^{-\frac{d}{2}}
	\frac{\Gamma(\frac{d}{2}+\nu_1)\Gamma(\frac{d}{2}+\nu_2)\Gamma(-\nu_1-\nu_2-\frac{d}{2})}
		{\Gamma(-\nu_1)\Gamma(-\nu_2)\Gamma(d+\nu_1+\nu_2)} (p^2)^{\frac{d}{2}+\nq+\nw}
\end{equation}
which shows that it scales proportional to the external momentum.
The left hand side of the ghost DSE, which consists only of the inverse dressed ghost 
propagator, is proportional to $(p^2)^{-\delta_{2,0}+1}$. The $1$ comes from the canonical 
dimension of the ghost propagator. The scaling dimensions on the right hand side are 
$d/2$ from the integral, $\delta_{0,2}-1$ from the gluon propagator, $\delta_{2,0}-1$ 
from the ghost propagator, $1/2$ from the bare ghost-gluon vertex, and $1/2$ 
from the dressed ghost-gluon vertex which features no anomalous scaling. This yields as 
condition for the IR exponents
\begin{align}
1-\delta_{2,0}&=\half{d}+\delta_{0,2}-1+\delta_{2,0}-1+\half{1}+\half{1} \, .
\end{align}
Defining the parameter $\kappa$ as $\ka := -\delta_{2,0}$, we recover 
in four dimensions the well-known result $\delta_{0,2}=2 \ka$. In $d$ dimensions we have \cite{Zwanziger:2001kw,Lerche:2002ep}
\begin{equation}
%%%%%begin
  %\delta_{2,0}=-\kappa \qquad , \qquad \delta_{0,2}=2\ka+1-\half{d} \, .
  \delta_{2,0}=-\kappa \qquad , \qquad \delta_{0,2}=2\ka+2-\half{d} \, .
%%%%%end
\end{equation}
A comparison between different dimensions via $\ka$ is not possible directly, because 
$\ka$ has different values for different $d$ \cite{Zwanziger:2001kw,Lerche:2002ep,Maas:2004se}. 
%%%%%begin
We discuss this point further below.
%%%%%end

In order to transform the infinite hierarchy of Dyson-Schwinger equations into a closed 
system we perform a skeleton expansion. This yields an infinite tower of graphs involving 
only primitively divergent vertices which we will analyze in a first step. The first order 
of the skeleton expansion of the DSE for the three-gluon vertex is depicted in 
\fref{fig:3g-DSE-skelexp-2}. We start with the ghost triangle which turns out to be one of the infrared leading diagrams. We have $d/2$ from the integral, $3(-\ka-1)$ from the three ghost propagators 
and $3/2$ from the three ghost-gluon vertices. We subtract the canonical dimension 
$1/2$ to get the anomalous IR exponent of the ghost triangle of the three-gluon 
vertex:
\equ{
\label{eq:3g}
\delta_{0,3}^{gh\Delta}=\half{d}+3(-\ka-1)+3\,\half{1}-\half{1}=-3\ka+\half{d}-2 \, .
}
%\fig{t}{fig:3g-DSE-skelexp-2}{./3g-DSE-skelexp,width=12cm}
%{The first order of the skeleton expansion of the DSE for the three-gluon vertex.} \\
\fig{t}{fig:3g-DSE-skelexp-2}{./3-gluon-DSE-skeleton,width=12cm}
{1-loop part of the skeleton expansion of the DSE for the three-gluon vertex.}
For the four-gluon vertex we can apply the same procedure. In fig. \ref{fig:4g-DSE-skelexp} we 
show the first order of its skeleton expansion. Here, also the ghost rectangle gives one of the IR 
dominant contributions. Simple counting of the powers yields for the IR exponent of 
this diagram
\equ{
\delta_{0,4}^{gh\square}=\half{d}+4(-\ka -1) +4 \,\halfo=-4\ka +\half{d}-2 \, .
}
%\fig{ht}{fig:4g-DSE-skelexp}{./4g-DSE-skelexp,width=12cm}
%{The first order of the skeleton expansion of the DSE for the four-gluon vertex.} \\
\fig{h}{fig:4g-DSE-skelexp}{./4-gluon-DSE-skeleton,width=15cm}
{1-loop part of the skeleton expansion of the DSE for the four-gluon vertex.}
Now we are in the position to derive the IR exponents of the other diagrams of the 
three-gluon vertex DSE in \fref{fig:3g-DSE-skelexp-2}. The gluon triangle contains three propagators as well as two dressed and one bare three-gluon vertex, so including the loop integral and subtracting the canonical dimension the IR exponent is $d/2+3(2\ka+1-d/2)+2(-3\ka +d/2-3/2)+1/2-1/2=0$. 
%Diagrams C, D and E contain a bare four-gluon vertex and have therefore an IR 
%exponent of $(-3\ka+\half{d}-2)+(4\ka-\half{d}+2)=\ka$. 
Correspondingly, the other 
corrections, including the two-loop diagrams that are not shown, are likewise subleading compared to the dominant ghost loop. 
%%%%%begin
%The same holds for the four-gluon vertex and the results are identical to eqs. (13) 
%and (15) in \cite{Alkofer:2004it} \textit{independent} of the dimension. 
The same holds for the four-gluon vertex. In general, the IR exponents of these 
purely gluonic diagrams are \textit{independent} of the dimension and therefore identical 
to eqs. (13) and (15) in \cite{Alkofer:2004it}.
%%%%%end 

With the scaling information of all primitively divergent Greens functions at hand we have 
all parts to calculate the IR exponent of an arbitrary $n$-point function. The different 
objects that can appear in corresponding general graphs are given in table \ref{tb:objects}.
\begin{table}[b]
\centering
\begin{tabular}{l|c|c|c|}
object & number of objects & scaling dimension \\ \hline
loop & l & $d/2$ \\
internal ghost line & $n_i$ & $\delta_{2,0}-1=-\ka -1$ \\
internal gluon line & $m_i$ & $\delta_{0,2}-1=2\ka+1-d/2$ \\
ghost-gluon vertex & $v_{2,1}$ & $1/2$ \\
bare 3-gluon vertex & $v^b_{0,3}$ & $1/2$ \\
dressed 3-gluon vertex & $v_{0,3}$ & $\delta_{0,3}+1/2=-3\ka+d/2-3/2$ \\
bare 4-gluon vertex & $v^b_{0,4}$ & $0$ \\
dressed 4-gluon vertex & $v_{0,4}$ & $\delta_{0,4}=-4\ka+d/2-2$ \\
\end{tabular}
\caption{Infrared behavior of the building blocks of general loop diagrams 
within the skeleton expansion.}
\label{tb:objects}
\end{table}
In terms of these building blocks the IR exponent of an arbitrary vertex $v$ is given by
\begin{align}\label{eq:ir-exp1}
\delta_v
	=&(l-m_i+v_{0,3}+v_{0,4})\half{d}+(2m_i-n_i-3v_{0,3}-4v_{0,4})\ka +\nnnl
	&\quad+\halfo(2m_i-2n_i+v_{2,1}+v^b_{0,3}-3v_{0,3}-4v_{0,4}-2v) \, ,
\end{align}
where $v$ is the canonical scaling dimension of the vertex.
This formula can be simplified using standard relations between the number of loops, the 
number of vertices and the number of propagators. 
The number of loops is given in terms of the internal lines and vertices via the topological 
relation
\equ{\label{eq:loop-number}
l=m_i+n_i+1-(v_{2,1}+v^b_{0,3}+v_{0,3}+v^b_{0,4}+v_{0,4}) \, .
}
Further topological relations that connect the number of lines of a given species with the 
corresponding vertices 
allow to rewrite the expression in terms of the number of external lines which are denoted 
by $m$ and $n$ for gluons and ghost-antighost-pairs, respectively:
\begin{align}
\label{eq:gluon-numbers}
&m+2m_i=v_{2,1}+3(v^b_{0,3}+v_{0,3})+4(v^b_{0,4}+v_{0,4}) \, , \\
\label{eq:ghost-numbers}
&n+n_i=v_{2,1} \, .
\end{align}
The canonical dimension of the vertex is finally given in terms of its external lines
\equ{\label{eq:vertex-dim}
v=\half{4-(2n+m)}=2-n-\half{m} \, .
}
Inserting eqs. (\ref{eq:loop-number}), (\ref{eq:gluon-numbers}), (\ref{eq:ghost-numbers}) 
and (\ref{eq:vertex-dim}) into eq. (\ref{eq:ir-exp1}) we get
\begin{align}\label{eq:ir-exp2}
\delta_{2n,m}&=(-n+1-v^b_{0,4}-v^b_{0,3})\half{d}+(4v^b_{0,4}+3v^b_{0,3}-m+n)\ka+\nonumber\\
		&\quad+(2v^b_{0,4}+2v^b_{0,3}+2n-2) \, .
\end{align}
This formula is still dependent on bare vertices. However, when considering different terms of 
the skeleton expansion of an $n$-point function, one easily sees that terms containing bare three- 
and four-gluon vertices are not the IR-dominant ones because contributions from these 
terms are
\begin{align}
&v^b_{0,3}(3\ka-\half{d}+2) \, , \\
&v^b_{0,4}(4\ka-\half{d}+2) \, ,
\end{align}
and thereby always add a positive power of momenta. 
%%%%%begin
%Correspondingly, we can neglect the number of bare three- and four gluon vertices and obtain 
Consequently, we can neglect these diagrams, when counting only IR leading ones, i.e. we can set $v^b_{0,3} = v^b_{0,4} = 0$ in (\ref{eq:ir-exp2}) and obtain 
%%%%%end
as a final result for the IR exponent of an $n$-point function the simple expression
\equ{\label{eq:ir-exp-dominant}
\delta_{2n,m}=(n-m)\ka+(1-n)\left(\half{d}-2\right) \, .
}
%%%%%begin
This relation verifies a posteriori the assumption discussed above fig.~\ref{fig:gh-DSE-again}:
the ghost-gluon scattering kernel is exactly as IR-divergent as to make the ghost-gluon
vertex finite in the IR, and this is independent of the value of the dimension $d$. 
%%%%%end

It remains to discuss higher order corrections in the skeleton expansion.
From \eref{eq:ir-exp-dominant} we can see immediately that there are infinitely many higher orders terms which have the same IR behavior as the leading term: Insertions which lead to higher orders only 
contribute with dressed vertices and propagators, and these were all considered in the 
derivation of \eref{eq:ir-exp-dominant}. These insertions 
%are shown below 
%\fref{fig:skeleton-insertions}. Alternatively one can check this directly by counting the 
with their corresponding IR exponents are given by: 
%(the $d/2$ stems from the additional loop integration 
%that is necessary when inserting these diagrams):
\begin{comment}
\fig{ht}{fig:skeleton-insertions}{./skeleton-insertions}
{Insertions that generate higher orders in the skeleton expansion.}
\end{comment}
\begin{center}
\begin{minipage}{2.5cm}
\epsfig{file=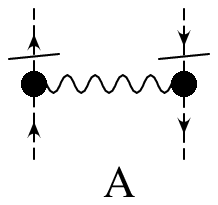}
\end{minipage}
\begin{minipage}{8cm}
$\; (2\ka+1-\frac{d}{2})+2(-\ka-1)+2\, \frac{1}{2}+\frac{d}{2}=0$
\end{minipage}
\end{center}
\begin{center}
\begin{minipage}{2.5cm}
\epsfig{file=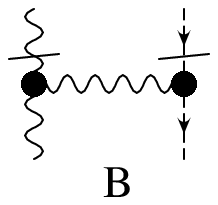}
\end{minipage}
\begin{minipage}{8cm}
$\; 2(2\ka+1-\half{d})+(-\ka-1)+(-3\ka+\half{d}-\half{3})+\halfo+\half{d}=0$
\end{minipage}
\end{center}
\begin{center}
\begin{minipage}{2.5cm}
\epsfig{file=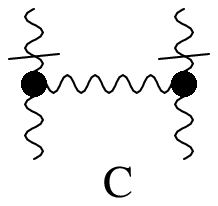}
\end{minipage}
\begin{minipage}{8cm}
$\; 3(2\ka+1-\half{d})+2(-3\ka+\half{d}-\half{3})+\half{d}=0$
\end{minipage}
\end{center}
\begin{center}
\begin{minipage}{2.5cm}
\quad \epsfig{file=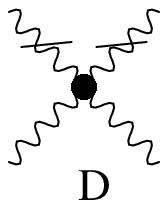}
\end{minipage}
\begin{minipage}{8cm}
$\; 2(2\ka+1-\half{d})+(-4\ka+\half{d}-2)+\half{d}=0$
\end{minipage}
\end{center}
\begin{comment}
\begin{itemize}
\item[A] \qquad $(2\ka+1-\frac{d}{2})+2(-\ka-1)+2\, \frac{1}{2}+\frac{d}{2}=0$
\item[B] \qquad $2(2\ka+1-\half{d})+(-\ka-1)+(-3\ka+\half{d}-\half{3})+\halfo+\half{d}=0$
\item[C] \qquad $3(2\ka+1-\half{d})+2(-3\ka+\half{d}-\half{3})+\half{d}=0$
\item[D] \qquad $2(2\ka+1-\half{d})+(-4\ka+\half{d}-2)+\half{d}=0$
\end{itemize}
\end{comment}
%We note that in contrast to a canonical scaling limit it is a non-trivial result that an 
%IR fixpoint compatible with the skeleton expansion exists in the discussed limit where 
%the theory is not conformal and involves 
%%%%%begin
% explicit mass scales.
%a dimensionful coupling.
%%%%%end
%It should be noted that in contrast to a canonical scaling limit we have kept the dimensionful coupling %$g$ fixed. Therefore, although the existence of power laws in the IR is expected, the usefulness of the %skeleton expansion to extract the IR exponents is a highly non-trivial result.
It should be noted that in contrast to
a uniform %%% a canonical
scaling limit we
have kept the
(dimensionful) %% %dimensionful
coupling $g$ fixed. The
IR power %%%infrared scaling
laws, eq. 
(\ref{eq:ir-exp-dominant}), are then valid for momenta $p^2 \ll g^{4/(4-d)}$.
The resulting momentum dependent couplings from the ghost-gluon vertex,
$\alpha^{gh}$, the three-gluon vertex, $\alpha^{3g}$, and the four gluon
vertex, $\alpha^{4g}$, (see \cite{Alkofer:2004it} for a definition) are
then given by
%\begin{equations}\label{eq:alpha_s}
 \begin{align}
 \alpha^{gh}(p^2) =& \frac{g^2}{4 \pi} \, [Z_{2,1}(p^2)]^2\,
%G^2(p^2) \, Z(p^2) \stackrel{\sim}{p^2 \rightarrow 0} (p^2)^{(4-d)/2} \,,
%G^2(p^2) \, Z(p^2) \; {\tiny \overrightarrow{p^2 \rightarrow 0}} \; (p^2)^{(4-d)/2} \,,
G^2(p^2) \, Z(p^2) \xrightarrow[p^2 \rightarrow 0] \; (p^2)^{(4-d)/2} \,,
\label{gh-gl} \\
 \alpha^{3g}(p^2) =& \frac{g^2}{4 \pi} \, [Z_{0,3}(p^2)]^2 \,
Z^3(p^2) \xrightarrow[p^2 \rightarrow 0] \; (p^2)^{(4-d)/2}\,,
 \label{3g} \\
 \alpha^{4g}(p^2) =& \frac{g^2}{4 \pi} \, [Z_{0,4}(p^2)] \,
 Z^2(p^2) \xrightarrow[p^2 \rightarrow 0] \; (p^2)^{(4-d)/2}\,. \label{4g}
\end{align}
%\end{equations}
Here $Z_{2,1}$ denotes the dressing function of the tree-level
structure of the ghost-gluon vertex and
$Z_{0,3}$ and $Z_{0,4}$  %%%$Z^{(0,3)},Z^{(0,4)}$
the
corresponding dressing functions of the three- and four-gluon vertices,
respectively. %%%
%%%and
To obtain these results
the scaling laws, eq. (\ref{eq:ir-exp-dominant}), have been used.
%It is a highly non-trivial result that the couplings universally scale
%with their canonical dimensions only. 
From this one may better understand
the usefulness of the skeleton expansion: apart from B all diagrammatic
pieces 
%in figure \ref{fig:skeleton-insertions} 
given above contain dressing factors
of the same multiplicity as the couplings, cf. eqs.  
%(\ref{eq:alpha_s}). 
(\ref{gh-gl}-\ref{4g}). Thus all
anomalous dimensions cancel and the canonical dimension $(4-d)/2$ of 
these couplings is cancelled %%% eaten up
by the canonical dimensions and the extra
loop. As a result the skeleton expansion indeed works independent of
the value of the spacetime dimension $d$ 
%This result suggests that it should likewise be possible to obtain such general counting rules from the
as could have been expected from the previously studied four-dimensional case \cite{vonSmekal:1997is,Zwanziger:2001kw,Lerche:2002ep,Alkofer:2004it,Fischer:2006vf}.

%%%%%%begin
\begin{figure}[t]
\centerline{\epsfig{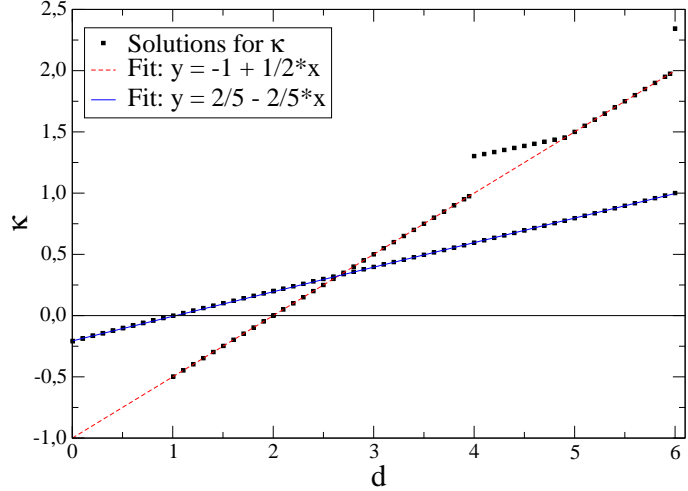}}
\caption{The (mostly two) possible solutions for the IR exponent $\kappa$ as a function
of dimension $d$, cf. also refs. \cite{Zwanziger:2003cf} and \cite{Maas:2004se}. The linear fits are exact in some ranges of dimension $d$ and approximate the real solutions to good precision in others. The first branch contains the solutions $\kappa=0.2$, $\kappa=0.3976...$ and $\kappa=0.5953...$ for $d=2,3$ and $4$, respectively, whereas the second one includes the solutions $\kappa=0$, $\kappa=0.5$ and $\kappa=1$ \cite{Zwanziger:2001kw}.}
\label{kappa}
\end{figure}
%%%%%%end

Finally, let us compare the IR behavior of the Yang-Mills Greens functions in 
different dimensions.  
%%%%%%begin
%In table \ref{tb:dim-dep} the complete momentum dependence of ghost 
%and gluon propagator and the three- and four-gluon vertices are shown. 
The corresponding values for the IR exponent $\kappa$ (using a bare ghost-gluon vertex) 
are displayed in fig.~\ref{kappa}. Apart from the region $d <1$ we always have two possible
values, which arise from an IR analysis of the ghost and gluon propagator DSEs. We show the IR exponents of the ghost and gluon propagators and the three- and four-gluon vertices in table \ref{tb:dim-dep}.
%We will show below, that the DSE for the three-gluon vertex excludes the branches
%in fig.~\ref{kappa} that starts at $\kappa=-1$ for $d=0$ and ends at $\kappa=2$ for $d=6$.
%Using the results of the other, physical branch, we show the IR exponents 
%of the ghost and gluon propagators and the three- and four-gluon vertices in table \ref{tb:dim-dep}.
%%%%%end
%Only the gluon propagator in 
%two dimensions could be finite for the second value of $\ka$, in contrast to the other cases 
%where it is infrared vanishing. 
%%%%%begin
%\begin{tabular}{l|c|c|c|c|c|c|c}
%Dimension &  & \multicolumn{2}{|c|}{$4$} & \multicolumn{2}{|c|}{$3$} & \multicolumn{2}{c}{$2$}\\ \hline
%$\ka$ & 			& $0.5953$ & $1$ & $0.3976$ &$0.5$	& $0.2$&$0$\\ \hline
%Ghost & $-\ka-1$ 		& $-1.6$ & $-2$	& $-1.4$&$-1.5$ 	& $-1.2$&$-1$\;\;\, \\ 
%Gluon & $2\ka+1-d/2$ 	& $0.2$ & $1$ & $0.3$&$0.5$ 	& $0.4$&$0$\\ 
%3-gluon & $-3\ka +d/2-3/2$ & $-1.3$ & $-1.5$ & $-1.2$&$-1.5$ 	& $-1.1$&$-0.5$ \\ 
%4-gluon & $-4\ka +d/2 -2$ 	& $-2.4$ & $-4$ & $-2.1$&$-2.5$ 	& $-1.8$&$-1$
%\end{tabular}
%\caption{The dimension dependence of the infrared behavior of Yang-Mills Green functions. 
%For each $d$ the two formally existing solutions of $\kappa$ \cite{Zwanziger:2001kw} are displayed.}
\begin{table}[b]
\centering
\begin{tabular}{c|c||c|c|c||c|c|c}
Dimension &   & {$4$} & {$3$} & {$2$} & {$4$} & {$3$} & {$2$} \\ \hline
$\ka$ & 	& $0.5953... \approx 0.6$  &  $0.3976... \approx 0.4$  & $0.2$ & $1$ & $0.5$ & $0$ \\ \hline
Ghost & $-\ka-1$ 		& $-1.6$    &  $-1.4$    & $-1.2$ & $-2$\;\;\, &$-1.5$ &$-1$ \\ 
Gluon & $2\ka+1-d/2$ 	        & \;\;\,$0.2$     &  \;\;\,$0.3$     & \;\;\,$0.4$ & $1$ & \;\;\,$0.5$ &$0$ \\ 
3-gluon & $-3\ka +d/2-3/2$      & $-1.3$    &  $-1.2$    & $-1.1$ & $-1.5$ & $-1.5$ & $-0.5$ \\ 
4-gluon & $-4\ka +d/2 -2$ 	& $-2.4$    &  $-2.1$    & $-1.8$ & $-4$\;\;\, & $-2.5$ & $-1$
\end{tabular}
\caption{The dimension dependence of the infrared behavior of Yang-Mills Green functions. 
For each $d$ the two solutions for $\kappa$ corresponding to the two alternative branches in fig. \ref{kappa} are displayed.}
%%%%%end
\label{tb:dim-dep}
\end{table}
\begin{comment}
\centering
\begin{tabular}{l|c|c|c|c}
Dimension &   & {$4$} & {$3$} & {$2$}\\ \hline
$\ka$ & 			& $0.5953... \approx 0.6$  &  $0.3976... \approx 0.4$  & $0.2$   \\ \hline
Ghost & $-\ka-1$ 		& $-1.6$    &  $-1.4$    & $-1.2$  \\ 
Gluon & $2\ka+1-d/2$ 	        & \;\;\,$0.2$     &  \;\;\,$0.3$     & \;\;\,$0.4$   \\ 
3-gluon & $-3\ka +d/2-3/2$      & $-1.3$    &  $-1.2$    & $-1.1$  \\ 
4-gluon & $-4\ka +d/2 -2$ 	& $-2.4$    &  $-2.1$    & $-1.8$
\end{tabular}
\caption{The dimension dependence of the infrared behavior of Yang-Mills Green functions. 
For each $d$ only the physical solution of $\kappa$ is displayed.}
%%%%%end
\label{tb:dim-dep}
\end{table}
\end{comment}
Whereas the situation in four dimensions is not conclusive, yet, the solutions belonging to the first branch in fig.~\ref{kappa} have been found in lattice studies in two and three dimensions. These results for the IR exponents of the propagators, the ghost-gluon vertex \cite{Cucchieri:2004sq} and the three-gluon vertex in two dimensions \cite{Maas:2007uv} agree within errors with the corresponding values given in table \ref{tb:dim-dep}. One can see that for this branch the qualitative behavior does not change in different dimensions. The other branch in fig.~\ref{kappa} that starts at $\kappa=-1$ for $d=0$ and ends at $\kappa=2$ for $d=6$ has not been seen in lattice simulations. We will give some additional arguments below that this branch may be unphysical.

\begin{figure}[ht]
\centering\epsfig{file=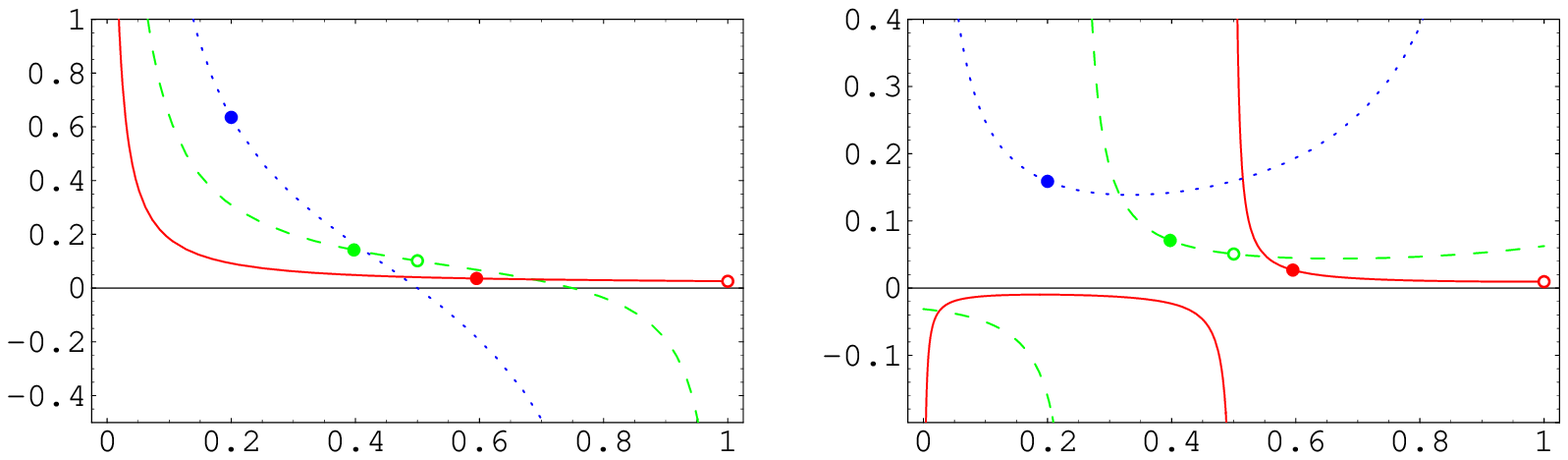,width=16cm}
\flushleft \vspace*{-3.2cm} $c_{2,0}$ \hspace*{7.2cm} $c_{0,2}$ \flushleft \vspace*{1.95cm} 
\hspace*{4.2cm} $\kappa$ \hspace*{7.55cm} $\kappa$
\caption{The $\kappa$-dependence of the unrenormalized dressing integrals appearing in the DSEs for the ghost (left) and gluon propagator (right). The dotted, dashed and solid lines show 
the curves for $d=2,3$ and $4$ respectively, see also ref. \cite{Lerche:2002ep}. The full and open dots represent the solutions of the first and second branch in fig. \ref{kappa}.}\label{fig:prop-kappa-dep}
\end{figure}

\paragraph*{Dependence of the loop integrals on the IR-exponent:}
The Dyson-Schwinger equations involve loop integrals over bare and dressed vertex functions. 
As just discussed the corresponding dressing functions exhibit a power law with 
appropriate infrared exponents. The corresponding coefficients depend on the IR exponent $\kappa$.  For the respective coefficients of the propagators, eq. (\ref{eq:power-laws-props}), which can be computed from the 2-point integral, eq. (\ref{eq:2-point-integral}), this $\kappa$-dependence is shown in fig. \ref{fig:prop-kappa-dep}, where $\kappa$-independent prefactors have been dropped.
%For comparison we also give the $\kappa$-dependence of 
%the corresponding 2-point function, eq. (\ref{eq:2-point-integral}), in the right panel 
%of fig. \ref{fig:kappa-dep} which shows basically the same qualitative behavior.
%with the exception of an additional pole at $\kappa=0$ in four dimensions.
The physical requirement that the dressing functions have to be positive restricts the possible values of $\kappa$ to $\kappa\leq0.5$ in two, $0.25<\kappa\leq0.75$ in three and $\kappa > 0.5$ in four dimensions \cite{Maas:2005rf}. We note that all known solutions fulfill these requirements.
The decisive point is, however, 
%Another interesting point is 
that although the curves in fig. \ref{fig:prop-kappa-dep} differ 
considerable, the physical values obtained from the Dyson-Schwinger solution are far away 
from these poles and thereby the $\kappa$-dependence in their vicinity is rather mild and 
qualitatively similar in each case. 

\begin{figure}[ht]
\centering\epsfig{file=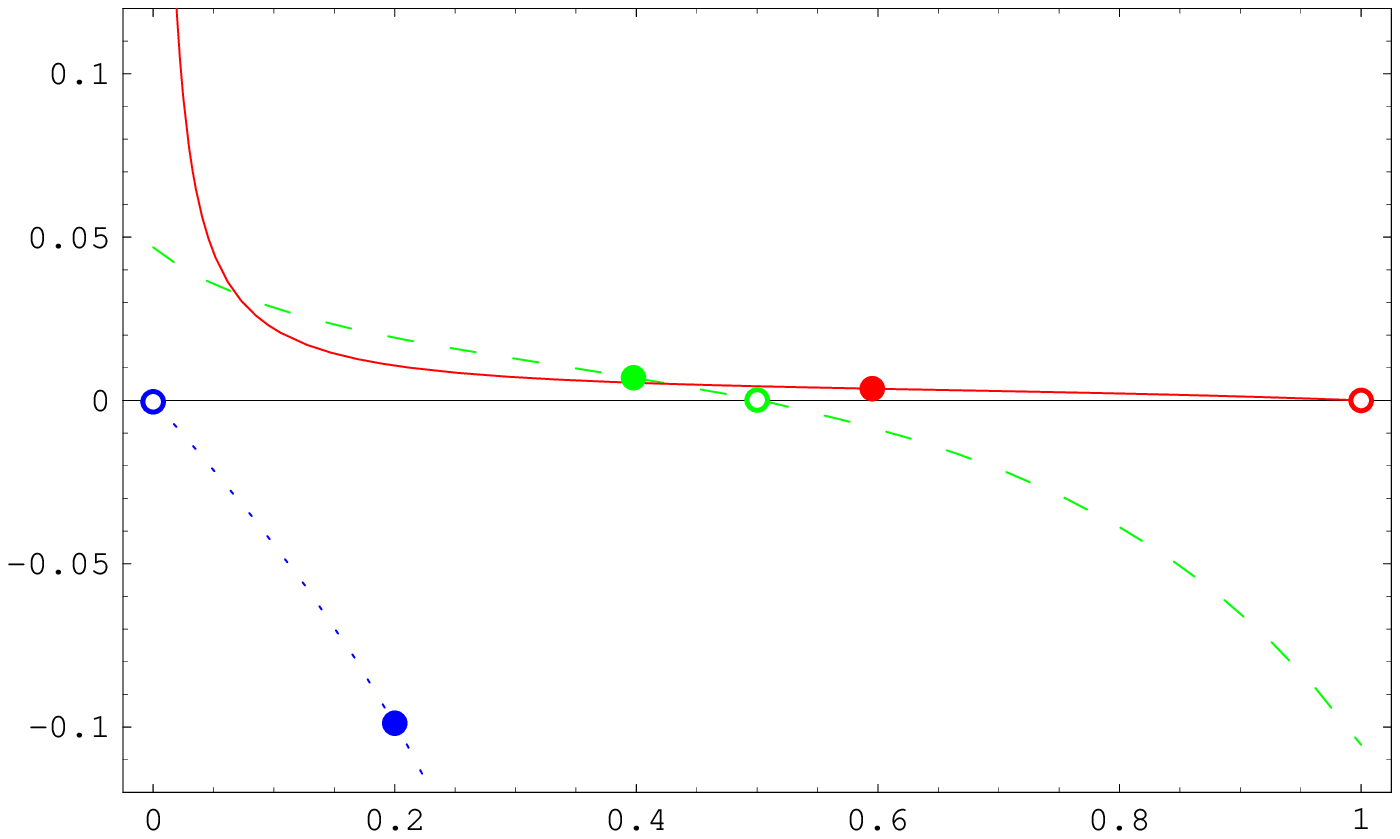,width=12cm}
\flushleft \vspace*{-4.55cm} \hspace*{1.4cm} $c_{0,3}$ \flushleft \vspace*{3.3cm} 
\hspace*{8.2cm} $\kappa$
\caption{The $\kappa$-dependence
%of the prefactor $c_{0,3}$ of the IR power-law
of the overlap of the unrenormalized ghost triangle with the tree-level tensor at the symmetric point. The dotted, dashed and solid lines show the curves for $d=2,3$ and $4$ respectively. The full and open dots represent the solutions of the first and second branch in fig. \ref{kappa}.}\label{fig:kappa-dep}
\end{figure}

In general the vertex integrals feature more 
complicated tensor structures with an increasing number of independent tensor components. 
However, there are general methods to decompose such tensor integrals to standard scalar 
integrals \cite{Davydychev:1991va}. The tensor integrals in the DSE for the ghost-gluon 
and three-gluon vertex reduce to 3-point integrals of the form
\begin{equation}
\label{3int}
  {I}_3 (p,q) \equiv \int \frac{d^d k}{(2\pi)^d} \frac{1}{((k+p)^2)^{\nu_1}} 
  \frac{1}{((k-q)^2)^{\nu_2}} \frac{1}{(k^2)^{\nu_3}}
\end{equation}
where any additional momenta due to vertex functions in the numerator are included 
in the denominators for appropriate values $\nu_i$.
A general expression for such scalar 1-loop 3-point integrals in arbitrary dimensions 
and with arbitrary
powers of the propagators has been obtained in \cite{Anastasiou:1999ui,Boos:1990rg}. 
An explicit evaluation of this result requires an expression for the resulting Appell 
function $F_4$ whose defining series does not converge in the considered Euclidean regime. 
A converging expression can be obtained by analytic continuation and has been given in \cite{Exton}. 
Unfortunately, the result given there is slightly incorrect. The correct and lengthy 
expression which agrees with a direct numerical integration of these integrals will be 
given in a forthcoming publication \cite{IR-vertices} where we will also give the detailed 
results for the different tensor structures and their kinematic dependence. Here we will 
merely discuss the dependence of the overlap of the tree-level tensor with the IR-dominant ghost triangle. 
%generic dependence of these integrals on the IR exponent $\kappa$. \\ 
%basic parameters of Yang-Mills theory.
%An important dynamical ingredient in the DSE solution is the 3-gluon vertex which has to 
%be implemented via a non-trivial ansatz even in the simplest approximation that ensures a 
%proper ultraviolet limit of the solution in four dimensions. 
According to the discussed power counting, the ghost loop presents the IR-leading contribution to the Dyson Schwinger equation for the three-gluon vertex in 
fig. \ref{fig:3g-DSE-skelexp-2}. Since this loop involves only ghost-gluon vertices that 
remain bare to leading order, its IR behavior can be analyzed semi-perturbatively using 
the scaling form of the dressed ghost propagators, see eq. (\ref{eq:power-laws-props}).
We consider the special kinematic configuration given by the symmetric point $p^2=q^2=r^2$. 
At this point the dressing function of the leading ghost loop correction to the three-gluon integral scales according to 
eq. (\ref{eq:3g}) as $Z_{0,3} = c_{0,3}(\kappa) \cdot (p^2)^{-3\kappa+d/2-2}$ where 
the dependence of the coefficient on the scaling parameter $\kappa$ has been made explicit. 
The coefficient is shown as a function of $\kappa$ in fig. \ref{fig:kappa-dep}, where the full and open points represent the values for the first and second branch in fig.~\ref{kappa} and $\kappa$-independent factors have been dropped again.
\begin{comment}
\begin{figure}[ht]
\centering\epsfig{file=./combined-kappa-dependence.eps,width=14cm}
\flushleft \vspace*{-2.5cm} $c_{0,3}$ \hspace*{6.15cm} $c_{0,2}$ \flushleft \vspace*{1.8cm} 
\hspace*{3.45cm} $\kappa$ \hspace*{6.45cm} $\kappa$
\caption{The $\kappa$-dependence of the scalar three point integrals (left) and the two 
point integrals (right) appearing in the ghost loop correction to the corresponding DSEs 
for the three-gluon vertex and the gluon propagator. The dotted, dashed and solid lines show 
the curves for $d=2,3$ and $4$ respectively.}\label{fig:kappa-dep}
\end{figure}
\end{comment}
 \\
The vertex integral has zeros at $\kappa=0$ in two, $\kappa=0.5$ in three and $\kappa=1$ in four dimensions which coincide precisely with the solutions of the second branch. These zeros depend on the kinematic configuration and appear at higher values of $\kappa$ away from the symmetric point.
(The corresponding scalar integrals, cf. eq. (\ref{3int}), feature even poles at these values of $\kappa$ which precisely cancel for the tree-level tensor.)
%found in \cite{Zwanziger:2001kw}.  
According to eq. (\ref{3g}) such zeros in the vertex dressing function would lead to an IR-vanishing coupling. Since Yang-Mills theory is apparently a strongly interacting theory, this indicates that these solutions might not be physically relevant. As in the case of the propagators the $\kappa$ dependence in the vicinity of the solutions of the first branch is similar in all dimensions. The negative value of the three-gluon vertex in two dimensions nevertheless leads to a positive IR limit of the coupling $\alpha^{3g}$ in eq. (\ref{3g}). However, it would cause problems in its renormalization group flow and might be an artifact of the skeleton expansion and cured by higher orders.
%In addition the ghost loop would not be the IR-leading graph in the DSE fig. \ref{fig:3g-DSE-skelexp-2} %and the 3-gluon vertex would be less singular than given in eq. \label{eq:3g}. Thereby the ghost 
%dynamics would dominate even more. 
\begin{comment}
%Since Yang-Mills theory is apparently not a non-interacting theory we can exclude the alternative 
%solutions $\kappa=0$ in two, $\kappa=0.5$ in three and $\kappa=1$ in four dimensions found 
%in \cite{Zwanziger:2001kw} 
%%%%%begin
%and displayed in fig.~\ref{kappa}.
%%%%%end
%This leaves us with one {\em unique} solution in each dimension. The same 
%conclusion was reached before in the case of four dimensions via a combined analysis of 
%both the DSE system and the functional RG \cite{Fischer:2006vf}.
The decisive point is, however, 
%Another interesting point is 
that although the curves in fig. \ref{fig:kappa-dep} differ 
considerable, the physical values obtained from the Dyson-Schwinger solution are far away 
from these poles and thereby the $\kappa$-dependence in their vicinity is rather mild and 
qualitatively similar in each case. 
\end{comment}

\paragraph*{Conclusions:}
We have studied generic features of the IR limit of $SU(N)$ Yang-Mills theory established 
in \cite{vonSmekal:1997is,Lerche:2002ep,Alkofer:2000wg,Alkofer:2004it} in more detail. As in the four-dimensional case, the IR behavior of Greens functions can be extracted via
a skeleton expansion in arbitrary dimensions. We find that the IR limit of Greens functions is surprisingly insensitive on the spacetime dimension. This already constitutes the first important result of this study. 
As a consequence Yang-Mills theory in lower dimensions has a 
qualitatively similar IR limit as the four dimensional theory. 
%%%%%begin
%Therefore, corresponding 
%lattice simulations should provide interesting qualitative information for the physical 
%case. In particular the confinement mechanism might possess identical features in different 
%dimensions. 
Thus the confinement mechanism might possess identical features in different 
dimensions. Corresponding lattice simulations should provide interesting qualitative 
information for the physical case. Indeed a recent study on large two dimensional 
lattices \cite{Maas:2007uv} finds infrared exponents in agreement with the value 
$\kappa=0.2$
%singled out by our analysis
, thus confirming the Gribov-Zwanziger
scenario also in two dimensions. 
%%%%%end

%The presented study of the $\kappa$-dependence of the integrals appearing in the DSEs allowed to 
%exclude one of the two known solutions in 2, 3 and 4 dimensions. Furthermore, 
The results 
on the mild $\kappa$-dependence of the remaining DSE solutions suggest that ghost dominance 
is a rather robust mechanism and should not depend on the details of the employed truncation 
scheme. This is further substantiated by the fact that the main non-linearities in the DSE 
system, which enable the non-trivial fixpoint, arise in the propagator equations whereas there 
is no non-linear feedback to leading order in the vertex equations.

\paragraph*{Acknowledgements:}
We are grateful to A. Maas and J. Pawlowski for valuable discussions and to A. Maas for a critical reading of the manuscript. This work has been supported in part by the DFG under contract AL279/5-1, by the FWF under contract M979-N16 and by the Helmholtz-University Young Investigator Grant VH-NG-332.

\end{document}